\documentclass[printer]{aa}
\usepackage{psfig}

\begin{document}

\title{Possible detection of hard X-ray afterglows of short 
$\gamma$-ray bursts}
\titlerunning{Afterglows of short GRBs}

\author{Davide Lazzati\inst{1}, Enrico Ramirez-Ruiz\inst{1}
\and Gabriele Ghisellini\inst{2}}
\authorrunning{Lazzati, Ramirez-Ruiz \& Ghisellini}

\offprints{D. Lazzati}

\institute{Institute of Astronomy, University of Cambridge, Madingley
Road, CB3 0HA Cambridge, England \\
\email{lazzati,enrico@ast.cam.ac.uk}
\and Osservatorio Astronomico di Brera, via Bianchi 46, I-23807 
Merate, Italy \\
\email{gabriele@merate.mi.astro.it}}

\date{}

\abstract{
We report the discovery of a transient and fading hard X-ray emission
in the BATSE lightcurves of a sample of short $\gamma$-ray bursts.  We
have summed each of the four channel BATSE light curves of 76 short
bursts to uncover the average overall temporal and spectral evolution
of a possible transient signal following the prompt flux.  We found an
excess emission peaking $\sim 30$~s after the prompt one, detectable
for $\approx 100$~s.  The soft power-law spectrum and the
time-evolution of this transient signal suggest that it is produced by
the deceleration of a relativistic expanding source, as predicted by
the afterglow model.
\keywords{Gamma rays: bursts -- Radiation mechanisms: non-thermal} }

\maketitle

\section{Introduction}
\label{sec:int}

Since their discovery in 1973 (Klebesadel et al. 1973), $\gamma$-ray
bursts (GRBs) have been known predominantly as brief, intense flashes
of high-energy radiation, despite intensive searches for transient
signals at other wavelengths.  Fortunately, the rapid follow-up of
{\it Beppo}SAX (Boella et al. 1997) positions, combined with
ground-based observations, has led to the detection of fading
emission in X-rays (Costa et al. 1997), optical (van Paradijs et
al. 1997) and radio (Frail et al. 1997) wavelengths. These afterglows
in turn enabled the measurement of redshifts (Metzger et al. 1997),
firmly establishing that GRBs are the most luminous known events in
the Universe and involve the highest source expansion velocities
(Piran 1999; M\'esz\'aros 2001).

Already in the early times of BATSE observations, it was noted that
the bursts can be divided into at least two classes based on their
duration and spectral hardness (Kouveliotou et al. 1993). The two
subclasses are usually classified as short and long GRBs, since the
difference in duration is more evident than in hardness (see
Fig.~\ref{fig:hr}). The long bursts, with a duration\footnote{$T_{90}$
is the time interval in which the observed fluence rises from $5\%$ to
$95\%$ of the total.} $T_{90}\ge2$~s, are usually characterized by
softer spectra with respect to the long ones. It is also generally
believed that the progenitor of the two classes of bursts may be
different (Lattimer \& Schramm 1976; Rees 1999): hypernovae (Paczynski
1998) or collapsars (MacFadyen \& Woosley 1999) for long bursts and
mergers of compact binary systems (Eichler et al.\ 1989) for short
ones.

The detection of afterglows that follow systematically long bursts has
been a major breakthrough in GRB science (van Paradijs et al. 2000).
Unfortunately no observation of this kind was possible for short
bursts.  Our physical understanding of their properties was therefore
put in abeyance, waiting for a new satellite better suited for their
prompt localization (Vanderspek et al. 1999).

In this letter we show that afterglow emission characterizes also the
class of short bursts. In a comparative analysis of the BATSE
lightcurves of short bursts (see \S~2), we detect a hard X-ray fading
signal following the prompt emission with a delay of $\sim 30$ s.  The
spectral and temporal behavior of this emission is consistent with the
one produced by a decelerating blast wave (see \S~3), providing a
direct confirmation of relativistic source expansion (Piran 1999).

\section{Data Analysis}

\begin{figure}
\psfig{file=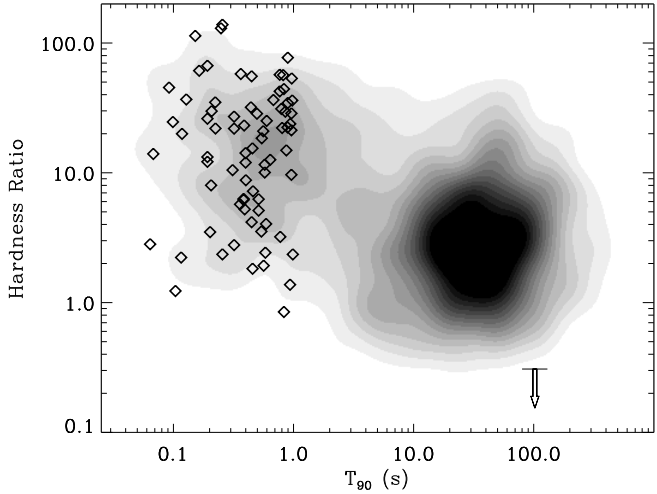,width=.48\textwidth}
\caption{
{The hardness-ratio vs. $T_{90}$ space distribution of BATSE GRBs.
The hardness ratio is computed as the ratio of the fluence of each
burst in the energy channels 3 and 4 divided by the fluence in the
channels 1 and 2.  Diamonds show the position in the diagram of the 76
short bursts selected for the search of an hard X-ray afterglow.  The
arrow in the lower right corner marks the upper limit on the hardness
ratio of the afterglow emission discussed in this paper.  }
\label{fig:hr}}
\end{figure}

The detection of slowly variable emission in BATSE lightcurves is a
non trivial issue, since BATSE is a non-imaging instrument and
background subtraction can not be easily performed. We selected from
the BATSE GRB catalog\footnote{The most recent version of the catalog
can be found at:
{\tt~http://www.batse.msfc.nasa.gov/batse/grb/catalog/}.}  (Paciesas
et al. 1999) a sample of short duration ($T_{90}\le1$~s), high
signal-to-noise ratio, GRB lightcurves with continuous data from
$\sim120$~s before the trigger to $\sim230$~s afterwards. We aligned
all the lightcurves to a common time reference in which the burst
(binned to a time resolution of 64~ms) peaked at $t=0$ and we binned
the lightcurves in time by a factor 250, giving a time resolution of
16.0~s. The time bin [$-8<t<8$~s] containing the prompt emission was
removed and the remaining background modelled with a $4^{\rm th}$
degree polynomial.  The bursts in which this fit yielded a reduced
$\chi^2$ larger than 2 in at least one of the four channels were
discarded. Note that we did not subtract this best fit background
curve from the data. This procedure was used only to reject
lightcurves with very rapid and unpredictable background fluctuations
and is based on the assumption that the excess burst or afterglow
emission is not detectable in a single lightcurve.  The position of
the selected bursts in the hardness duration plane is shown with
diamonds in Fig.~\ref{fig:hr}. This procedure yielded a final sample
of 76 lightcurves, characterized by an average duration
$\langle{T}_{90}\rangle=0.44$~s and fluence
$\langle{\cal{F}}\rangle=2.6\times10^{-6}$~erg~cm$^{-2}$.

\begin{figure*}
\centerline{\psfig{file=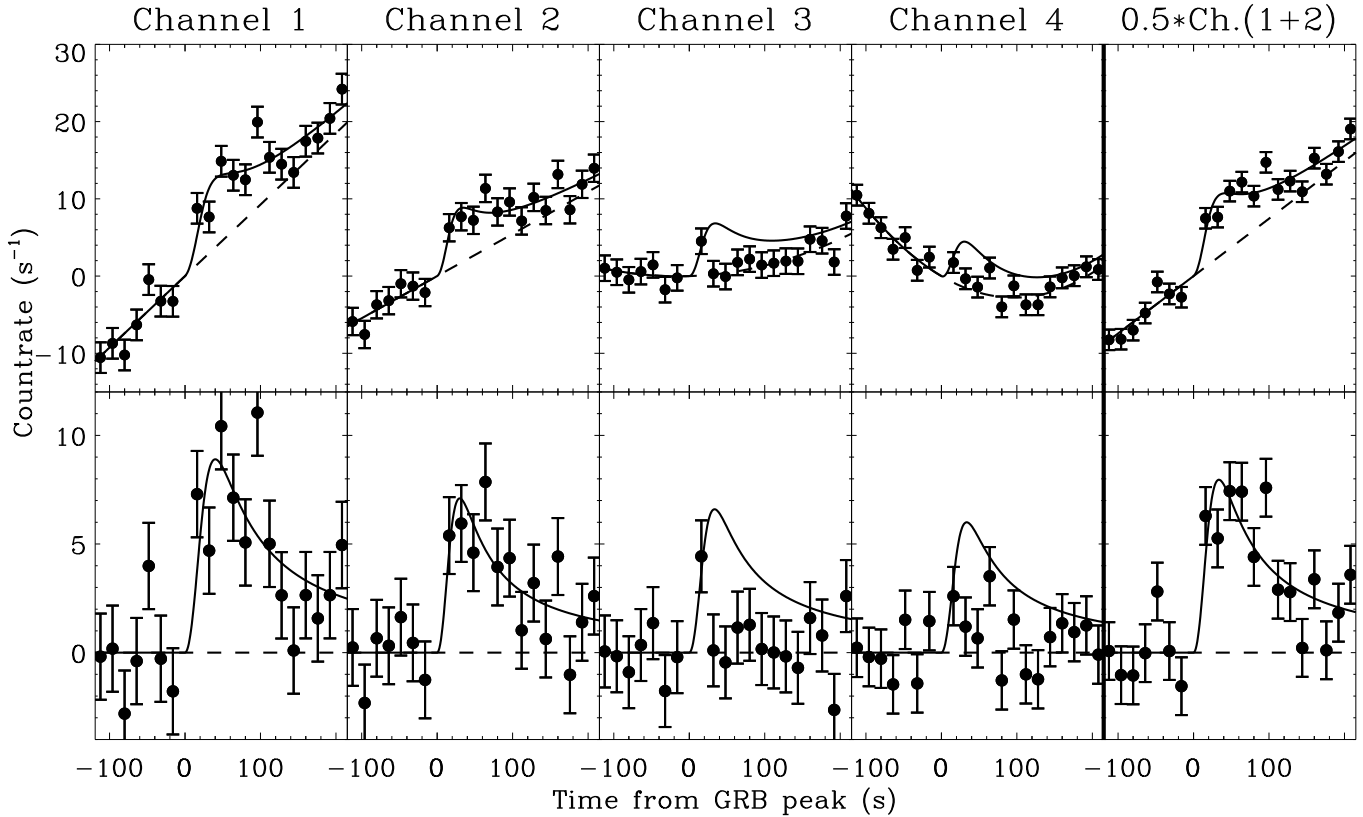,width=\textwidth}}
\caption{{Overall lightcurves in the 4 BATSE channels (from left to
right) of the sample of short bursts (see text). The rightmost panels
show the average signal in the first and second channels. The time
interval of the burst emission has been excluded. The upper panels
show the lightcurves without background subtraction (a constant has
been subtracted in all panels for viewing purposes in order to have
zero counts at $t=0$). The solid line is the best fit background plus
afterglow model (in the channel 3 and 4 panels the $3\sigma$ upper
limit afterglow is shown). The dashed line shows the background
contribution in all channels. The lower panels show the same data and
fit after background subtraction.}
\label{fig:lcur}}
\end{figure*}

To search for excess emission following the prompt burst, we added the
selected binned lightcurves in the four channels independently.  The
resulting lightcurves are shown in the upper panels of
Fig.~\ref{fig:lcur} by the solid points. Error bars are computed by
propagating the Poisson uncertainties of the individual lightcurves.

The lightcurves in the third and fourth channels can be successfully
fitted with polynomials. The third (110--325~keV) and fourth
($>325$~keV) channel lightcurves can be fitted with a quadratic model,
yielding $\chi^2/{\rm d.o.f.}=17/18$ and $\chi^2/{\rm d.o.f.}=18.5/17$
respectively.  In the first two channels, a polynomial model alone
does not give a good description of the data. In the first
(25--60~keV) channel, a cubic fit yields $\chi^2/{\rm d.o.f.}=42/16$,
while in the second (60--110~keV) we obtain
$\chi^2/{\rm d.o.f.}=26/16$.  A more accurate modelling of the first
two channels lightcurves can be achieved by allowing for an afterglow
emission following the prompt burst.  We model the afterglow
lightcurve with a smoothly joined broken power-law function:
\begin{equation}
L_A(t) = {{3\,L_A}\over{\left({{t_A}\over{t}}\right)^2 + {{2\,t}\over{t_A}}}}; 
\;\;\; t>0
\label{eq:ag}
\end{equation}
which rises as $t^2$ up to a maximum $L_A$ that is reached at time
$t_A$ and then decays as $t^{-1}$ (Sari 1997). Adding this afterglow
component to the fit, we obtain $\chi^2/{\rm d.o.f.}=28.5/16$ and
$14.7/16$ in the first and second channels, respectively\footnote{This
lightcurve is appropriate for the bolometric afterglow luminosity and
it is only an approximation for the flux integrated in a spectral
band.}. The $\chi^2$ variation, according to the F-test, is
significant to the $\sim3.5\sigma$ level in both channels. The fact
that the fit in the first channel is only marginally acceptable should
not surprise. This is because the excess is due to many afterglow
components peaking at different times, and has therefore a more
``symmetric'' shape than Eq.~\ref{eq:ag}. A fully acceptable fit can
be obtained with a different shape of the excess, but we used the
afterglow function for simplicity. By adding together the first two
channels, the afterglow component is significant at the $4.2\sigma$
level. The results of the fit are reported in Tab.~\ref{tab:fit}.

A similar attempt to detect long timescale emission in BATSE
lightcurves was performed by Connaughton (2000).  In that paper excess
emission after the main burst was detected in long GRBs, but no signal
was detected for short GRBs.  The main difference with respect to the
present work is that Connaughton (2000) performed the analysis on the
total lightcurves (summed over the four channels) and that the short
bursts selection was based only on the duration ($T_{90}<2$~s) and not
on the signal-to-noise ratio.  In addition, Connaughton (2000)
performed a burst by burst background subtraction measuring the
background of the single events by averaging the signal of the 15 {\it
Compton}GRO orbits before and after the burst.  Since these data are
not available in the BATSE public archive, we cannot perform the same
analysis and check our results with this method.  However, by adding
the four channel data, the significance of our detection drops to less
than $2\sigma$ while, relaxing the constraint on the signal-to-noise
ratio in the sample selection, we detect no afterglow in the sum of
the resulting 280 lightcurves sample.  We believe that a check of the
robustness of this detection against the background subtraction
technique of Connaughton (2000) would yield interesting results.

\begin{table}
\begin{tabular}{l|c|c|c}
 & E (keV) & $L_A$ (cts~s$^{-1}$) & $t_A$ (s) \\ \hline \hline
Channel \#1 & [25-60] & $8.9\pm2.6$ & $40\pm16$ \\
Channel \#2 & [60-110] & $7.1\pm2.4$ & $30^{+16}_{-10}$ \\
Channel \#1+2 & [25-110] & $16\pm3.5$ & $33.5^{+24}_{-15}$ \\
Channel \#3 & [110-325] & $<6.6$ & $33.5$ (fixed) \\
Channel \#4 & $>325$ & $<6.0$ & $33.5$ (fixed)
\end{tabular}
\caption{{Fit results. Quoted errors at $90\%$ levels, upper limits 
at $3\sigma$ level.}
\label{tab:fit}}
\end{table}

\subsection{Further statistical tests}

\begin{figure*}
\centerline{\psfig{file=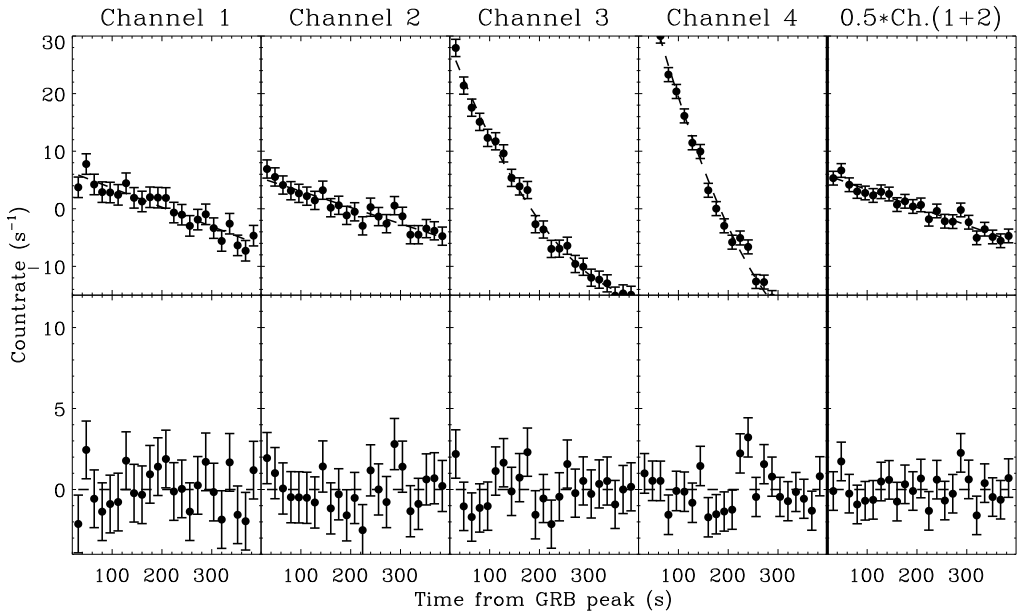,width=0.75\textwidth}}
\caption{{Same as Fig.~\ref{fig:lcur} but for a sample of blank BATSE 
lightcurves (see text). No excess is present in these lightcurves.
The only marginally significant deviation from a polynomial fit is in
the $4^{\rm th}$ channel, where a parabolic fit (showed with a dashed
line) yields $\chi^2/{\rm d.o.f.}=24/20$.}
\label{fig:blank}}
\end{figure*}

To test the robustness of the result we reanalyzed the data with the
following procedure. We fit the background of each lightcurve in the
time intervals $-120<t<-8$~s and $\tilde t<t<230$~s with a $4^{\rm
th}$ degree polynomial with $50<\tilde t<120$~s. We then subtracted
the best fit background from each lightcurve at all times. We finally
averaged the residuals in the four BATSE channels. Even though the
actual shape of the residuals depends on the value of $\tilde t$, in
all cases we found excesses in the first two BATSE channels and a null
results in the third and fourth channels. We checked for the
possibility of a single burst lightcurve dominating this detection by
recursively subtracting one lightcurve. Performing the same analysis
the statistical significances given above were not altered. As a
further test we applied the same analysis to a sample of low
signal-to-noise short GRBs and to a sample of blank BATSE background
lightcurves\footnote{These lightcurves were obtained extracting the
data in the interval $50<t<400$ ($t=0$ is set by the GRB trigger) from
a sample of 62 short low signal-to-noise GRB lightcurves.}
(see Fig.~\ref{fig:blank}). 
In both cases we obtained a null result.

In all the analysis and tests described above, we assumed that the
uncertainty in the countrates are purely Poissonian. To check the
correctness of this assumption we computed the errors of the mean
lightcurve from the dispersion of the sample and not by propagation of
the statistical error of each curve. Even though the actual result
depends slightly on the degree of the polynomial function used to
detrend each individual lightcurve, we find that the errors shown in
Fig.~\ref{fig:lcur} are increased by less than $10\%$ and the
significance of the feature is only marginally modified.

We finally explored the possibility that the detected signal is
residual emission from the burst by fitting a background plus a single
power-law to the data. These fits did not significantly decrease the
$\chi^2$ with respect to background alone (see also \S~3 for the
spectral analysis).  

\section{Discussion}

\begin{figure}
\psfig{file=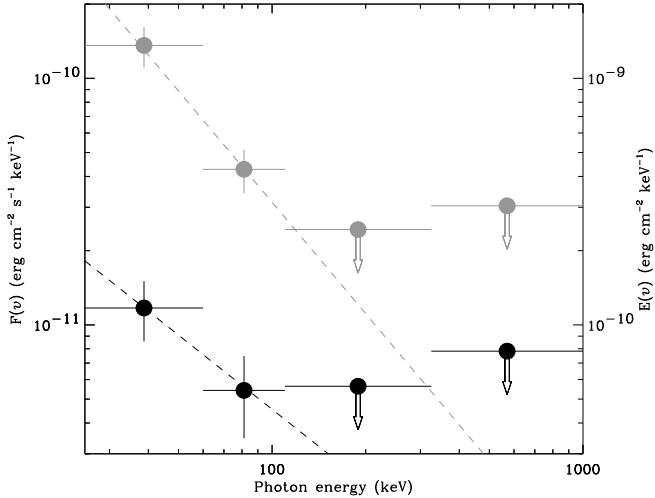,width=.48\textwidth}
\caption{{Spectrum of the peak afterglow emission. Black dots (and left 
vertical axis) show the spectrum at $t=30$~s. Gray dots (right
vertical axis) show the time integrated spectrum as obtained from the
four BATSE channel counts. Error bars are for $90\%$ uncertainties,
while arrows are $3\sigma$ upper limits.}
\label{fig:spec}}
\end{figure}

In order to understand whether the excess is residual prompt burst
emission or afterglow emission, we computed its four channel
spectrum. To convert BATSE count rates to fluxes, we computed an
average response matrix for our burst sample by averaging the matrices
of single bursts obtained from the {\tt~discsc\_bfits} and
{\tt~discsc\_drm} datasets. The resulting spectrum is shown in
Fig.~\ref{fig:spec}. The dark points show the spectrum at $t=30$~s
(the peak of the afterglow in the second BATSE channel), which is
consistent with a single power-law $F(\nu)\propto\nu^{-1}$ (black
dashed line). Grey points show the time integrated spectrum (fluences
has been measured with a growth curve technique), consistent with a
steeper power-law $F(\nu)\propto\nu^{-1.5}$ (grey dashed line).  These
power-law spectra are much softer than any observed burst spectrum
(independent of their duration). The upper limit of the hardness ratio
is shown in Fig.~\ref{fig:hr} with an arrow in the lower right
corner. This spectral diversity, together with the fact that a single
power-law does not fit the data, suggest that the emission is not due
to a tail of burst emission but more likely to an early hard X-ray
afterglow. This also confirms earlier predictions that the mechanism
responsible for the afterglow emission is different from that of the
prompt radiation (e.g. Piran 1999).

Hard X-ray prompt afterglows have been detected for the long bursts
GRB~920723 (Burenin et al. 1999) observed by GRANAT/SIGMA and
GRB~980923 (Giblin et al. 1999) observed with BATSE. In both these
events a power-law decaying tail was observed at the end of the
prompt emission associated with a spectral transition. The derived
$t_A$ in these bursts was $\sim 6$~s and $9.6$~s, respectively. The
observed peak time of bolometric afterglow emission is given by (Sari
1997):
\begin{equation}
t_A = 93.6\,(1+z)\, \left({E \over 10^{52} {\rm erg}}\right)^{1/3}\,
\left( {\Gamma\over 100}\right)^{-8/3}\,n^{-1/3} 
\;\;{\rm s}
\label{eq:ta}
\end{equation}
where $z$ is the redshift, $E$ is the isotropic equivalent kinetic
energy of the fireball, $\Gamma$ is the asymptotic Lorentz factor and
$n$ the baryon number density of the external medium.

The shorter $t_A$ detected in long bursts may hence reflect a higher
Lorentz factor $\Gamma$ or external density $n$ in long GRBs with
respect to short ones (see Eq.~\ref{eq:ta}). A smaller external
density would indeed be required if short bursts are associated to
mergers of compact objects (Eichler et al. 1989).  Less likely, given
the constraints on the $\langle V/V_{\rm max}\rangle$ of the two
classes, the longer peak time may be due to a higher redshift or a
larger kinetic energy of the fireball (Lee \& Petrosian 1997; Schmidt
2001).  On the other hand, the optical flash observed in the long
burst GRB990123 (Akerlof et al. 1999), had a peak time $t_A\sim40$~s,
showing that a dispersion of the $\Gamma$ and $n$ parameters is
present within a single burst class.

The detected peak flux of
$\sim10^{-11}$~erg~cm$^{-2}$~s$^{-1}$~keV$^{-1}$ ($\sim0.004$ mJy) at
50~keV can be compared with sparse measurements of short/intermediate
GRBs. Measurements at comparable times ($10-100$s) have been possible
only with robotic optical telescopes for three short bursts detected
by BATSE (Kehoe et al. 2001). These measurements yielded upper limits
of $\sim5$~mJy in unfiltered CCD images. Assuming a flat
$F(\nu)\propto\nu^0$ spectrum this upper limits are much brighter than
our detected average afterglow. An optical and radio afterglow was
detected and intensely observed for the intermediate duration
($T_{90}\sim2$~s) burst GRB000301C (Masetti et al. 2000). The
multiwavelength afterglow of this burst can be fitted by an external
shock model (Panaitescu \& Kumar 2001b). The extrapolated flux at
50~keV $\sim50$~seconds after the burst event is $\sim 0.01$~mJy, in
good agreement with our detection.

Another interesting comparison can be made with the early afterglow of
long GRBs.  Connaughton (2000) finds that on average the BATSE
countrate is $\sim150$~cts~s$^{-1}$ at $t=50$~s after the main
event. In our short GRB sample, the countrate at the same time is
$\sim15$~cts~s$^{-1}$.  Since the luminosity of the average early
X-ray afterglow is representative of the total isotropic energy of the
fireball (Kumar 2000), we can conclude that the isotropic equivalent
energy of the short bursts is on average ten times smaller than that
of the long ones (or that their true energy is the same, but the jet
opening angle is three times larger). Indeed, the $\gamma$-ray fluence
of long bursts is on average ten times larger than that of short
bursts.

In the case of short bursts, the analysis of the afterglow emission is
made easier by the lack of superposition with the prompt burst
flux. For this reason the time and luminosity of the afterglow peak
can be directly measured while in long bursts it had to be inferred
from the shape of the decay law at longer times (Burenin et al. 1999;
Giblin et al. 1999).  In our case, however, the lightcurve in
Fig.~\ref{fig:lcur} is the result of the sum of many afterglow
lightcurves, with different peak times and luminosities. For a given
isotropic equivalent energy $E$, afterglows peaking earlier
(with larger $\Gamma$) are expected to be brighter and should dominate
the composite lightcurve. On the other hand, for a given Lorentz
factor $\Gamma$, afterglow peaking earlier (with lower $E$) are
dimmer. The fact that the $\sim35$~s timescale is preserved, suggests
that there are only few very energetic bursts with a large bulk
Lorentz factor.

In the external shock model, the multiwavelength lightcurve of
the afterglow can be calculated (see Panaitescu \& Kumar 2000 and
references therein), given the properties of the fireball (initial
bulk Lorentz factor $\Gamma$ and isotropic equivalent energy $E$), the
external medium density $n$ and the shock properties (the
equipartition parameters $\epsilon_e$ and $\epsilon_B$, the slope of
the electron power-law distribution $p$ and the $\gamma$-ray
efficiency $\eta$).  Some of the parameters in the above set are
poorly constrained, like the equipartition parameters $\epsilon_B$ and
$\epsilon_E$. In the following we will adopt the logarithmically
averaged $\langle\epsilon_B\rangle=0.004$ and
$\langle\epsilon_e\rangle=0.06$, obtained by Panaitescu \& Kumar
(2001) by fitting lightcurves and spectra of long GRB afterglows.

The peak in the lightcurve observed at $t\sim 35$~s can be associated
either (a) with the beginning of the afterglow (fireball deceleration,
see Eq.~\ref{eq:ta}) or (b) with the transition of the peak of the
spectrum in the observed band. In the first case, the peak time
coincides with the deceleration time. In the second case, the peak
time follows the deceleration time with some delay and the peak of the
spectrum must be at $\sim 50$~keV at $t\sim35$~s. These two
constraints ($t_a<50$~s and $25<h\nu_{\rm peak}<110$~keV) can be
simultaneously satisfied only for $\Gamma>200$ and
$\epsilon_e>0.1$. In addition, the observed spectrum is expected to be
either $F(\nu)\propto\nu^{-1/2}$ (fast cooling) or
$F(\nu)\propto\nu^{-(p-1)/2}$ (slow cooling, where $p\sim2.5$ is the
index of the power-law distribution of electrons), harder than what
is observed.

If the observed peak is instead due to the time of deceleration of the
fireball [case (a) above], a consistent solution can be found for the
parameter set $z\sim0.4$, $n\sim10^{-1}$~cm$^{-3}$, $\Gamma=100$,
$E\sim5\times10^{51}$~erg and $\eta=0.2$. An electron distribution
$n(\gamma)\propto\gamma^{-2.5}$ was also assumed, consistent with the
spectrum of Fig.~\ref{fig:spec} (in this regime the spectral slope is
given by $F(\nu)\propto\nu^{-p/2}$). The data are hence consistent
with the possibility that short bursts have a similar energy budget to
the long ones, but with a $3-10$ times larger jet opening angle.  On
this basis, it is also possible to predict the flux and fluence at the
time at which follow-up observations are typically performed.  In
X-rays, usually observed at $t=8$~h, the predicted flux is
$F_{[2-10]{\rm keV}}\sim3\times10^{-13}$~erg~cm$^{-2}$~s$^{-1}$.  In
the optical, at $t=12$~h, we predict a magnitude $R\sim22$.  Both
these fluxes should be within the detection capabilities of present
instrumentation (see also Panaitescu et al. 2001).

It is interesting also to explore whether a consistent solution can be
found for very small redshifts, or even for short bursts exploding in
our own galaxy. This is plausible for the following parameter set:
$d=100$~kpc; $n\sim10^{-4}$~cm$^{-3}$; $\Gamma=20$ and
$E\sim2\times10^{43}$~erg.  However in this case the typical frequency
emitted by the injected electrons is in the optical band (Panaitescu
\& Kumar 2000) and the solution is only valid if the power-law of
accelerated electrons extends for more than four orders of magnitude.
In addition, the analysis of the number counts of short and long
bursts suggest a similar average redshift for the two populations
(Schmidt 2001).

We must also remain aware of other possibilities.  For instance, we
may be wrong in assuming that the central object goes dormant after
producing the initial explosion. A sudden burst followed by a slowly
decaying energy input could arise if the newly formed black hole
slowly swallows the orbiting torus around it or if the central object
becomes a rapidly-spinning pulsar rather than a black hole (Rees
1999).  This luminosity may dominate the continuum afterglow at early
times before the blast wave decelerates.  Under this interpretation,
the hard X-ray transient following the prompt emission could be
attributed to the central object itself rather than to a standard
decelerating blast wave. Contrary to what is observed, this emission
should smoothly decay after the main episode, unless this energy is
converted into a relativistic outflow which is in turn converted to
radiation at a larger radius.

\begin{acknowledgements}
We thank the anonymous referee for his carefully and constructive
reading of the manuscript. We thank Luigi Stella for many stimulating
comments and suggestions and Martin J. Rees, Sergio Campana, William
Lee and Elena Rossi for useful conversations.
\end{acknowledgements}

\end{document}